\documentclass[reprint, superscriptaddress, aps, prb, floatfix]{revtex4-2}

\usepackage{amsmath}
\usepackage{amssymb}

\usepackage{graphicx}

\usepackage[colorlinks=true]{hyperref}
\hypersetup{
    colorlinks = true,
    linkcolor = blue,
    citecolor = blue,
    urlcolor = blue
}

\begin{document}
\title{Universal Magnetic Structure Prediction from Atomic Coordinates with Near-Experimental Accuracy}

\author{Abhijatmedhi Chotrattanapituk}
\thanks{These authors contributed equally to this work.}
\affiliation{Quantum Measurement Group, MIT, Cambridge, MA 02139, USA}
\affiliation{Department of Electrical Engineering and Computer Science, MIT, Cambridge, MA 02139, USA}

\author{Ryotaro Okabe}
\thanks{These authors contributed equally to this work.}
\affiliation{Quantum Measurement Group, MIT, Cambridge, MA 02139, USA}
\affiliation{Department of Chemistry, MIT, Cambridge, MA 02139, USA}

\author{Eunbi Rha}
\affiliation{Quantum Measurement Group, MIT, Cambridge, MA 02139, USA}
\affiliation{Department of Nuclear Science and Engineering, MIT, Cambridge, MA 02139, USA}

\author{Mariya Al-Hinai}
\affiliation{Department of Electrical Engineering and Computer Science, MIT, Cambridge, MA 02139, USA}

\author{Eugene Jiang}
\affiliation{Department of Electrical Engineering and Computer Science, MIT, Cambridge, MA 02139, USA}
\affiliation{Department of Physics, MIT, Cambridge, MA 02139, USA}

\author{Daniel Pajerowski}
\affiliation{Neutron Scattering Division, Oak Ridge National Laboratory, Oak Ridge, TN 37831, USA}

\author{Yongqiang Cheng}
\affiliation{Neutron Scattering Division, Oak Ridge National Laboratory, Oak Ridge, TN 37831, USA}

\author{Joshua J Turner}
\affiliation{SLAC National Accelerator Laboratory, Menlo Park, CA 94025, USA}

\author{Mingda Li}
\email{mingda@mit.edu}
\affiliation{Quantum Measurement Group, MIT, Cambridge, MA 02139, USA}
\affiliation{Department of Nuclear Science and Engineering, MIT, Cambridge, MA 02139, USA}

\begin{abstract}
    Magnetic order is a fundamental property of materials, governing collective behavior and enabling a broad range of functionalities. Yet magnetic structure remains difficult to determine: experiments are costly and specialized, while first-principles methods often struggle with the noncollinear and incommensurate orders found in real materials. Here we introduce magnetic structure network (MSN), an $E(3)$ equivariant graph neural network that predicts both collinear and non-collinear magnetic structures directly from atomic crystal structures, trained directly on experimentally determined structures from MAGNDATA. By proposing the primitive modulated structure representation (PMSR), we are able to encode commensurate and incommensurate structures in a unified way without symmetry assumptions. The model achieves strong performance across all modulation components and reconstructs experimental magnetic structures with high fidelity. Our approach provides a scalable framework for rapid magnetic structure prediction and opens a route to data-driven discovery of magnetic materials. 
\end{abstract}

\maketitle
\section{Introduction}
Magnetic structure is one of the most fundamental properties in condensed matter, exposing the interplay of symmetry, correlation, topology, and entanglement while underpinning technologies ranging from magnetic storage and spintronics to electric motors, medical imaging, and quantum information science. Despite the importance, determining magnetic structures has been a long-standing challenge. Experimental techniques like neutron diffraction are widely regarded as the gold standard for magnetic structure determination, but they are resource-intensive and require large, high-quality samples and specialized expertise; even powder diffraction may be insufficient to fully resolve magnetic order, while single-crystal measurements pose an even greater challenge \cite{wollan1948diff,lovesey1984neutron}. Resonant elastic X-ray scattering (REXS) can, in principle, provide equivalent information, but its analysis becomes highly complex due to the underlying probing mechanisms, rendering REXS generally complementary to neutron diffraction rather than a standalone solution \cite{fink2013resonant}. Other experimental techniques, including M\"ossbauer spectroscopy \cite{yoshida2013mossbauer}, muon spin rotation ($\mu$SR) \cite{le2011muon}, nuclear magnetic resonance (NMR) \cite{hore2015nuclear}, X-ray magnetic circular dichroism (XMCD) \cite{stohr1999exploring,van2014x}, and Lorentz microscopy \cite{grundy1968lorentz}, among many others, can provide partial or indirect information about magnetic structures but are insufficient to determine the full magnetic configuration on their own.

On the computational side, density functional theory (DFT) and its extensions (DFT+U, hybrid functionals) rely on prior assumptions about spin configurations and become computationally prohibitive for complex non-collinear orders or incommensurate structures, which require large supercells \cite{cohen2012challenges, zhang2021high}. While effective lattice models, such as the Heisenberg or Hubbard Hamiltonians \cite{marston1989large}, can capture strong correlations, accurately mapping models to materials or deriving model parameters from first principles remains a bottleneck. Furthermore, advanced many-body solvers face severe fundamental constraints: exact diagonalization (ED) \cite{lin1990exact} is restricted to small clusters, density matrix renormalization group (DMRG) \cite{jiang2008dmrg} and tensor networks are most effective in one- and low-dimensional problems, quantum Monte Carlo (QMC) \cite{sandvik1991quantum} suffers from the fermion sign problem, and dynamical mean-field theory (DMFT) \cite{kvashnin2015exchange} struggles with long-range spatial correlations. In fact, all these methods incur computational costs far exceeding those of DFT. Therefore, a fast, general-purpose approach to predict magnetic structures approaching at experimental accuracy is urgently needed.

Recent machine learning approaches have begun to address aspects of magnetic structure prediction. Random forest models have been used to estimate spin and orbital magnetic moments and classify magnetic ordering in uranium- and neptunium-based compounds with 76\% accuracy, highlighting physically meaningful descriptors such as $f$-subshell occupation and structural parameters \cite{ghosh2020ml}. For two-dimensional materials, SISSO-based models achieved a 90\% success rate in distinguishing ferromagnetic from antiferromagnetic ordering, emphasizing the role of $d$-orbital occupancy, symmetry, and spin–orbit coupling \cite{acosta2022ml}. More recently, equivariant neural networks trained on the MAGNDATA database classified magnetic order and predicted propagation vectors, correctly identifying magnetic structures over 70\% of the time and non-magnetic materials with 90\% reliability \cite{merker2022ml}. Subsequently, gradient-boosted models pushed classification and propagation-vector prediction beyond 90\% accuracy, extending this analysis to large-scale computational datasets \cite{fahmy2025ml}. However, these approaches remain limited to property prediction or partial characterization of magnetic order, rather than full magnetic structure determination.

To address these limitations, we introduce magnetic structure network (MSN), an $E(3)$-equivariant graph neural network that predicts complete magnetic structures directly from atomic coordinates without requiring prior symmetry annotations or predefined spin configurations. Trained on over 2,300 experimentally verified structures from the MAGNDATA database \cite{gallego2016magndata}, the predictive capability of MSN is achieved through two critical ingredients. First, we introduce the primitive modulated structure representation (PMSR), which fundamentally unifies the treatment of commensurate and incommensurate structures by encoding all magnetic orders into a shared mathematical space of propagation vectors, moment amplitudes, and phase shifts. This enables the unified training of both commensurate and incommensurate magnetic structures, significantly enhancing data efficiency. Second, we design a novel dual-level output architecture: the model simultaneously predicts global propagation vectors as continuous graph-level properties and site-specific Fourier components as node-level features. By embedding these capabilities within an equivariant architecture that strictly preserves crystallographic symmetries, MSN ensures a physically consistent mapping from local atomic environments to extended magnetic modulations.

\section{Results}
\subsection{Primitive Modulated Structure Representation}
\begin{figure*}[!htb]
    \includegraphics[width = 0.8\linewidth]{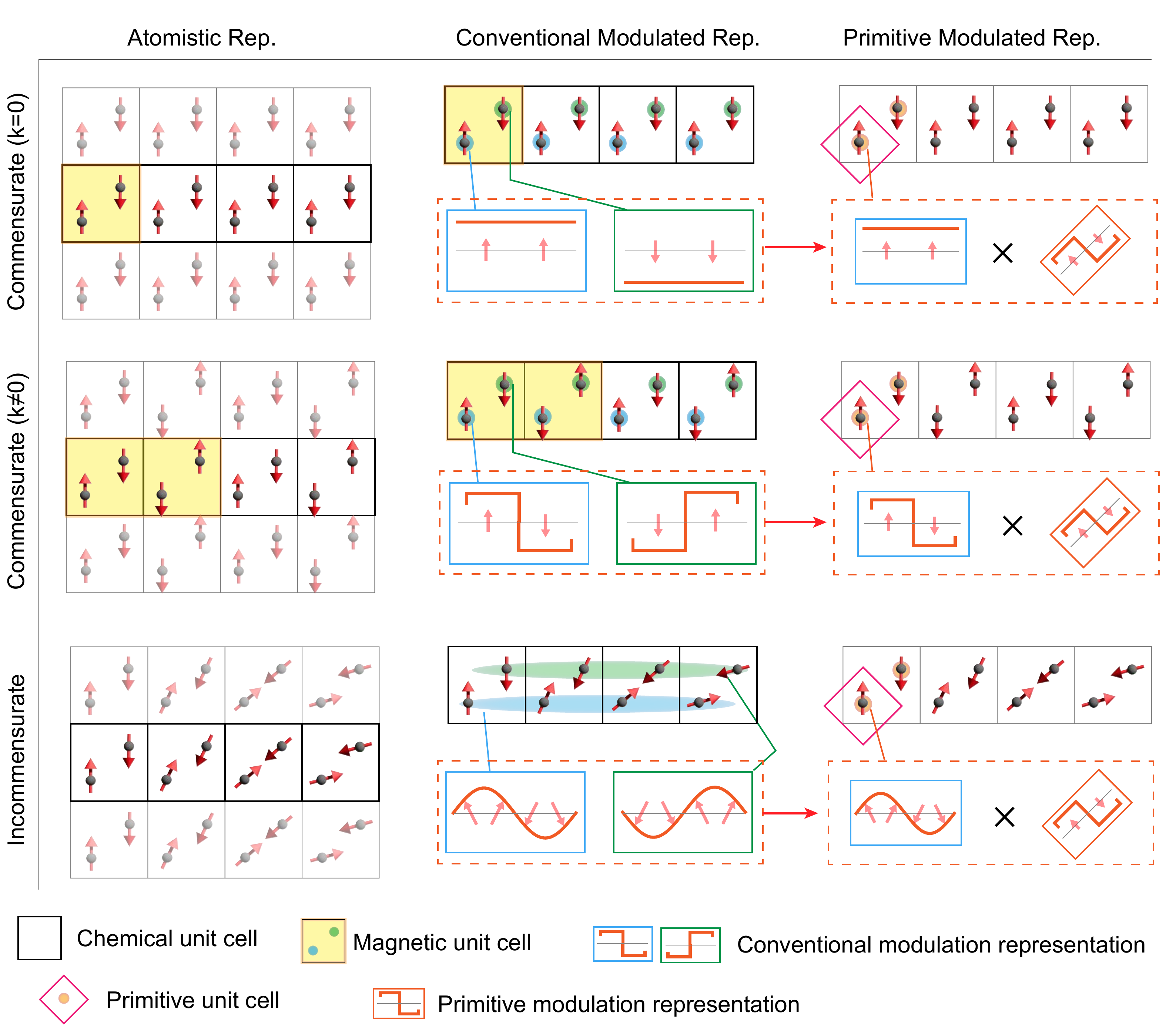}
    \caption{Primitive Modulated Structure Representation (PMSR). Commensurate structures both with (middle row) and without (top row) non-zero propagation vectors ($\mathbf{k}$) can be represented as periodic structures called magnetic unit cells that are rational multiples of their chemical unit cells. In contrast, by definition, there is no number of chemical unit cell that can capture the periodicity of incommensurate structures (bottom row). The modulated structure representation can captures the essences of all magnetic structure classes by, on top of embedding atomic sites in a unit cell with the magnetic moment vectors ($\mathbf{m}$), providing how the magnetic moment vectors change (get modulated) when one move to another unit cells. PMSR (right column) is a modulated structure representation whose modulation basis is the primitive chemical unit cell of the material. One can directly achieve PMSR from atomic representation (left column), magnetic unit cell representation or modulated structure representations with non-primitive unit cell basis (middle column) through some combinations of continuous/discrete Fourier transforms and Fourier mode subtractions/additions.
    }
    \label{fig:PMSR}
\end{figure*}

In commensurate magnetic structures, the periodicity of the magnetic structure is a rational multiple of that of the atomic structure. Starting from the primitive chemical (atomic) unit cell, one can therefore construct a finite super cell in which the magnetic moments repeat. This super cell, referred to as the magnetic unit cell, serves as a unit cell for both the atomic and magnetic structures. Within this representation, symmetry-equivalent atomic sites in the primitive unit cell may carry distinct magnetic moments, making the description analogous to a conventional crystal structure with magnetic moment vectors attached to each site as shown in figure~\ref{fig:PMSR} where the structure repeat itself with periodicity of the magnetic unit cell represented in yellow-shaded area. However, this representation becomes impractical for incommensurate structures or commensurate structures with large periodicity denominators, since it would require an infinitely large or prohibitively large super cell. It is also inconvenient for machine learning applications, as the number of magnetic moments to predict depends on the magnetic periodicity.

To address these limitations, we instead represent magnetic structures in terms of site-resolved modulation functions defined on chemical unit cell. For each atomic site, the magnetic moment is described not only by its local value but also by its periodic modulation across unit cells. Since periodic magnetic order can generally be expressed as a sum of sinusoidal components, the natural representation is a Fourier decomposition of the magnetic moments over reciprocal space. In this form, the magnetic structure is specified by a set of propagation vectors and corresponding complex Fourier components associated with each atomic site with site-specific magnetic moments determined by
\begin{equation}\label{eq:modulation_site}
    \mathbf{m}_{\alpha,\mathbf{n}} = \sum_\mathbf{k}\widetilde{\mathbf{m}}_{\alpha,\mathbf{k}}\exp(2\pi i\mathbf{k}\mathbf{n}^\top)\,
\end{equation}
where $\mathbf{m}_{\alpha,\mathbf{n}}$ is the magnetic moment of site $\alpha$ in the $\mathbf{n}$ unit cell, $\mathbf{k}$ is propagation vector (in reciprocal lattice vector unit) that modulate the magnetic moment from the $\mathbf{0}$ unit cell, and $\widetilde{\mathbf{m}}_{\alpha,\mathbf{k}}$ is the modulation amplitude of the corresponding $\mathbf{k}$ mode. For real-valued magnetic moments, the Fourier components additionally satisfy the conjugate symmetry relation between opposite propagation vectors. In other words, we can write
\begin{equation}\label{eq:modulation_site_cc}
    \mathbf{m}_{\alpha,\mathbf{n}} = \sum_\mathbf{k}\widetilde{\mathbf{m}}_{\alpha,\mathbf{k}}\exp(2\pi i\mathbf{k}\mathbf{n}^\top) + \text{c.c.}
\end{equation}
where the summation only run over half of the $\mathbf{k}$-space compare to Eq.~(\ref{eq:modulation_site}), and c.c. represents complex conjugate of the first term.

Since commensurate and incommensurate structures can be consider as modulation of magnetic moments of the $\mathbf{0}$ unit cell with different rationality of the propagation vector components, this representation can describe both commensurate and incommensurate magnetic orders from within a chemical unit cell as shown in the middle column of Fig.~\ref{fig:PMSR}. In fact, all incommensurate structure in MAGNDATA database are stored in the format that describe magnetic moment modulation from some chemical unit cell. 

To standardize all magnetic structures for machine learning, we transform every representation into a primitive modulated structure representation (PMSR), where the primitive chemical unit cell is always used as the modulation basis. This transformation consists of expressing all site coordinates, propagation vectors, and magnetic Fourier components in the primitive lattice basis. When the original representation (either magnetic unit cell for commensurate structure or the chemical unit cell for incommensurate structure) uses a larger unit cell, multiple sites may map onto the same primitive atomic position but with distinct Fourier components. This indicates an additional commensurate modulation of the Fourier components themselves. To resolve this, we identify the periodicity of this secondary modulation, perform a discrete Fourier decomposition, and combine the resulting components with the original propagation vectors. Wave vectors outside the chosen reciprocal-domain convention are mapped back with the appropriate phase corrections, and an additional site-dependent phase shift is applied so that all modulation functions are centered at their corresponding atomic sites.

The resulting PMSR (right column in Fig.~\ref{fig:PMSR}) provides a unified and compact representation capable of describing commensurate and incommensurate magnetic structures from within a fixed primitive unit cell. This standardization removes ambiguity in representation, enables direct comparison between different magnetic structures, and provides a consistent target format for machine learning models. With PMSR, given the material structure, we can directly get the atomic sites with in each primitive chemical unit cell. Hence, the only missing parts to describe any magnetic structure are propagation vectors as well as their corresponding Fourier components which will be the main prediction targets of our model.

\subsection{Magnetic Structure Networks}
\begin{figure*}[!htb]
    \includegraphics[width = 0.9\linewidth]{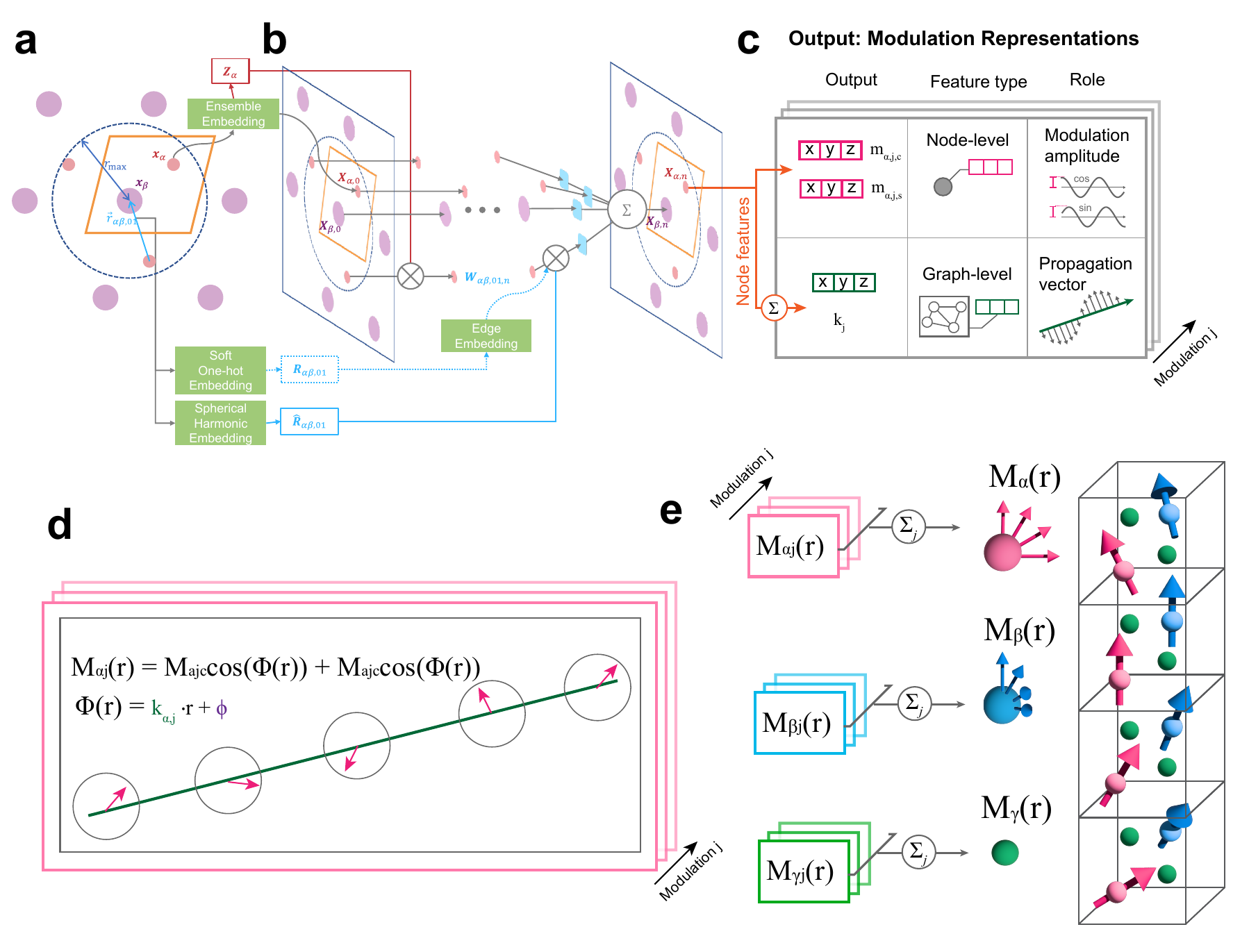}
    \caption{Magnetic structure network (MSN). \textbf{a)} Construction of periodic graph of each material structure with atoms in a primitive chemical unit cell as nodes and edges defined only between neighboring atoms with interatomic distance under certain cut off radius $\mathbf{r}_\text{max}$. \textbf{b)} $E(3)$-equivariant message passing with connections according to graph in \textbf{a}. \textbf{c)} Outputs of the model which can be divided into node level and graph level outputs which require extra aggregation from all nodes at the last layer of the model. The prediction of magnetic propagation vectors which are the properties of the whole structure uses the graph level output while the prediction of site-specific Fourier components uses the node level output. \textbf{d)} The modulation of magnetic moment across unit cells due to each propagation vector. \textbf{e)} The magnetic structure reconstruction from all propagation vector components.
    }
    \label{fig:MSN}
\end{figure*}

We developed the magnetic structure network (MSN), an $E(3)$-equivariant convolutional crystal graph neural network, to predict magnetic structures directly from atomic structures. Owing to crystal periodicity, we can construct a finite graph for any crystalline material from its primitive chemical unit cell by representing each atomic site in the cell as nodes and encoding neighboring interactions as edges where the interactions that go outside the cell are mapped back to the cell through periodic boundary conditions as shown in Fig.~\ref{fig:MSN}a and the top part of Fig.~\ref{fig:MSN}b.

To ensure computational tractability, edges are also defined only between pairs of atoms separated by less than a fixed cutoff radius $r_\text{max}$, allowing the graph to capture local chemical environments while remaining translationally invariant through the use of relative atomic displacements. Our model can also accommodate doped materials and partially occupied sites as overlapping nodes with associated occupancies, which are incorporated explicitly into the graph connectivity.

The model inputs consist of three main components: atomic features, interatomic displacements, and reciprocal lattice basis vectors. Atomic features serve both as initial node embeddings and fixed node attributes during message passing. Interatomic displacement information is split into angular and radial components where angular dependence is represented by spherical harmonic expansions of bond directions, while interatomic distances are embedded using soft one-hot encodings followed by a learnable multilayer perceptron to generate radial features. To supplement the finite graph with global information about crystal periodicity, the reciprocal lattice basis vectors are provided directly as vector-valued inputs. This is particularly important for predicting magnetic propagation vectors, whose associated modulation wavelengths may extend beyond the graph cutoff radius. With the provided inputs, our neural network operates through $E(3)$-equivariant message passing where node features are iteratively updated by aggregating information from neighboring nodes as shown in Fig.~\ref{fig:MSN}b.

Since our prediction targets contain both site specific information like magnetic moment Fourier components as well as structure-wide information like propagation vectors, our models need to predict both levels of information as well. For the prediction of the magnetic propagation vectors, we treated it as graph-level quantities obtained by aggregating model output information from all nodes. The prediction is  subdivided into the predictions of the number of non-zero propagation vectors, and the sequence of propagation vectors itself of length equal to the predicted number of non-zero propagation vectors.

On the other hand, the model predicts magnetic moment Fourier components for individual atomic sites, i.e., the node level output from the last layer of the model can be directly used. Because non-magnetic atoms significantly outnumber magnetic ones, a preliminary binary classification is used to identify magnetic atoms before attempting moment prediction, preventing the model from collapsing toward trivial zero-moment solutions. For magnetic sites, i.e., sites that passed the filter by magnetic atoms classifier, the Fourier components are predicted directly as pseudo-vector outputs. Figure~\ref{fig:MSN}c, summarize all outputs from the models.

Once we obtain the propagation vectors and Fourier components, we can directly reconstruct the magnetic structure following the Eq.~(\ref{eq:modulation_site_cc}) as shown in Fig.~\ref{fig:MSN}c-d. The specific representation of each prediction output, the target loss functions used, as well as the reconstruction process will be explain in greater detail in the method section.

\subsection{Magnetic Structure Reconstruction}
\begin{figure*}[!htb]
    \includegraphics[width=\linewidth]{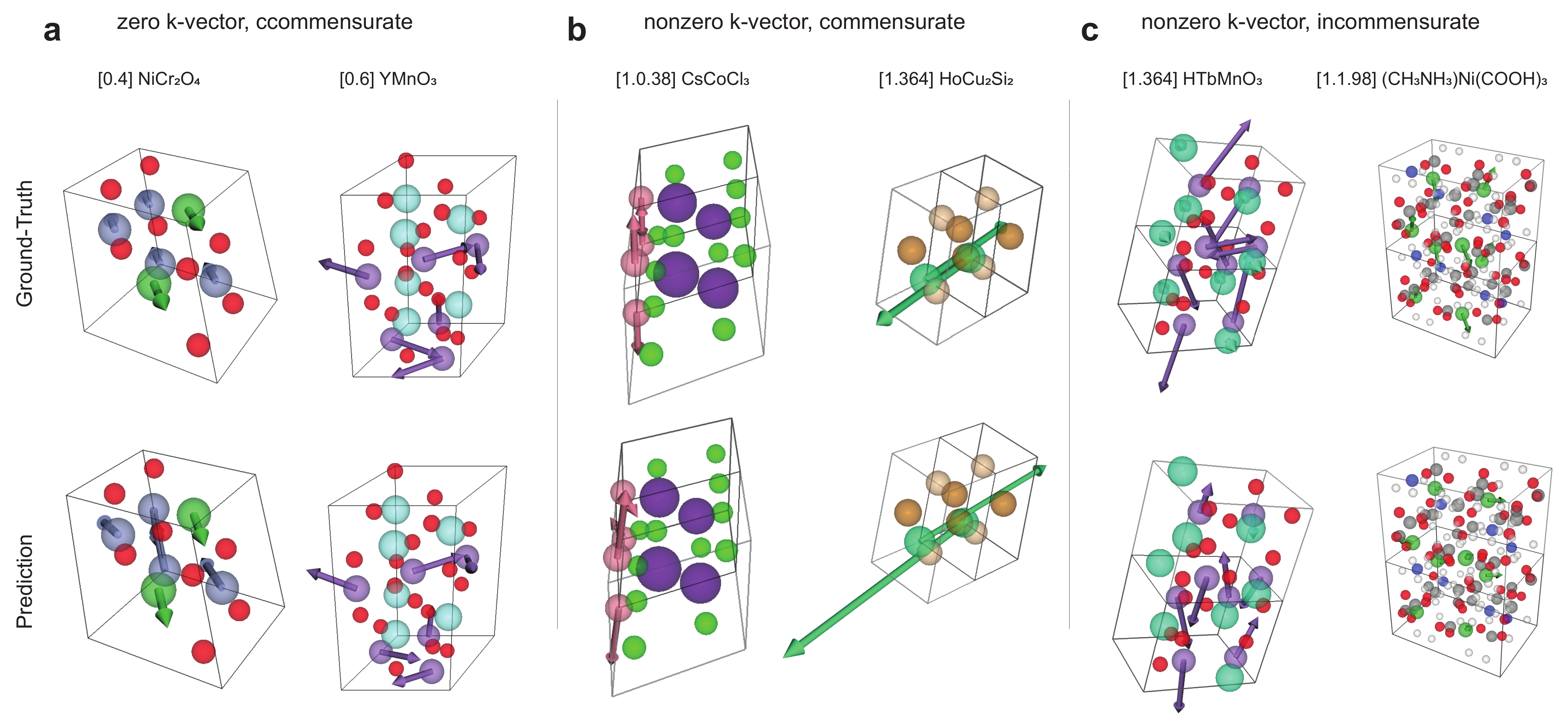}
    \caption{Predicted magnetic structures. For each material (MAGNDATA ID, formula), the ground-truth structure
    from MAGNDATA (top) is shown alongside the model prediction (bottom).
    \textbf{(a)} Commensurate structures with zero propagation vector ($\mathbf{k} = \mathbf{0}$).
    \textbf{(b)} Commensurate structures with a non-zero rational propagation vector. The visualization supercell is constructed from the ground-truth $\mathbf{k}$-vector
    (expanding up to a factor of 2 per lattice direction) to make the spatial modulation
    of magnetic moments visible.
    \textbf{(c)} Incommensurate structures: the propagation vector has an irrational component, and the same supercell convention is applied for visualization.
    Moment directions are shown as arrows on each magnetic site, and arrow lengths are proportional to the predicted moment amplitude.}
    \label{fig:mag_out}
\end{figure*}

To qualitatively evaluate the model's ability to reconstruct full magnetic structures, we present representative predictions spanning the three major structural classes in MAGNDATA. Figure~\ref{fig:mag_out} shows pairs of ground-truth (GT) and predicted magnetic structures for (a) commensurate structures with zero propagation vector ($\mathbf{k}=\mathbf{0}$, i.e.\ purely ferromagnetic or simple antiferromagnetic order), (b) commensurate structures with a non-zero rational propagation vector, and (c) an incommensurate structure. Ground-truth structures are taken directly from the MAGNDATA database. For the non-zero propagation vector cases, we use the ground-truth $\mathbf{k}$-vector to determine an appropriate supercell for visualization (expanding up to a factor of 2 along each lattice direction), so that the spatial modulation of the magnetic order is visible. The predicted structures are reconstructed end-to-end from the model's outputs following Eq.~\ref{eq:modulation_site_cc}: the propagation vectors and complex Fourier components are predicted independently, and the real-space moments are assembled from those components. Across all three classes, the predicted moment orientations and modulation patterns are in good qualitative agreement with the ground truth, demonstrating that our framework and the equivariant architecture together enable the model to capture the distinct character of each magnetic order type without structure-class-specific supervision.

\section{Discussion}
Our proposed magnetic structure network (MSN) provides a scalable framework for predicting full magnetic structures directly from atomic crystal structures, a capability enabled by the combination of primitive modulated structure representation (PMSR) and E(3)-equivariant graph neural networks. By representing magnetic order through propagation vectors and site-resolved Fourier components, MSN unifies the prediction of commensurate and incommensurate magnetic structures within a fixed and physically meaningful output space. Beyond magnetic structure reconstruction, this framework may be extended to other challenging magnetic prediction tasks, such as estimating magnetic excitation spectra, spin Hamiltonian parameters, or temperature-dependent magnetic phase transitions. The efficiency of MSN also enables a new paradigm for magnetic materials discovery, where approximate magnetic ground states can be rapidly predicted across large materials databases, providing guidance for experimental characterization and improved initial conditions for first-principles calculations.

Our investigation also reveals several limitations of the current model. First, the relatively small size of the MAGNDATA dataset may limit generalization, particularly for rare or highly complex magnetic phases. Second, the current framework adopts a Heisenberg-type description and therefore does not explicitly model anisotropic spin interactions, despite the E(3)-equivariant architecture being in principle capable of representing such behavior. This assumption primarily arises from limitations in systematically identifying anisotropic contributions across the available experimental dataset. In addition, the Fourier-component representation is not unique: global phase freedom and global spin rotation can produce physically equivalent magnetic structures, particularly for incommensurate systems. As a result, additional phase and rotational alignment are required during training and evaluation, increasing computational complexity.

These limitations may be addressed through both improved data availability and further model development. Expanding experimentally verified magnetic structure datasets, potentially combined with pretraining on computationally generated magnetic configurations, would improve model robustness and generalization. Incorporating explicit anisotropic interactions and alternative invariant representations could further reduce ambiguity in Fourier-component prediction while extending the framework beyond the present Heisenberg-type assumption. More broadly, the ability to integrate physical symmetry directly into the learning framework highlights the potential of MSN as a physically interpretable machine-learning approach. Together, MSN opens the possibility for automated magnetic structure determination and high-throughput exploration of magnetic materials, advancing data-driven discovery in complex magnetic systems.

% Our investigation also reveals several limitations of the current model. First, the relatively small size of the MAGNDATA dataset may restrict generalization, particularly for rare or highly complex magnetic phases. Second, the current framework assumes Heisenberg-type magnetic structures and therefore does not explicitly account for anisotropic spin interactions or more complex magnetic order parameters. In addition, the non-uniqueness of Fourier component representations requires additional phase and rotational alignment during training, increasing computational complexity and potentially limiting scalability for structures with many modulation components.

% These limitations may be addressed through both improved data availability and further model development. Expanding experimentally verified magnetic structure datasets would improve model robustness, while incorporating additional physical constraints or alternative invariant representations could reduce ambiguity in Fourier-component prediction. More broadly, the ability to integrate physical symmetry directly into the learning framework highlights the potential of MSN as a physically interpretable machine-learning approach. Together, MSN opens the possibility for automated magnetic structure determination and high-throughput exploration of magnetic materials, advancing data-driven discovery in complex magnetic systems.

\section{Methods}
\subsection{Data Preparation and Preprocessing}
We trained our models on MAGNDATA database\cite{gallego2016magndata}, a database for experimentally observed magnetic crystal structure materials in the Bilbao crystallographic server. There are, at the time of this work, 2,348 published magnetic structure with downloadable structure files in Magnetic Crystallographic Information File (MCIF) format. Each entry in the database includes magnetic unit cell lattice parameters, magnetic space group symmetries, symmetrized atomic compositions and positions, and magnetic moment vectors for commensurate structure. For incommensurate one, the magnetic space group is replaced by magnetic super space group, and the magnetic moments is supplemented with the incommensurate Fourier components. 

While commensurate magnetic material files can be almost perfectly parsed by \verb|Pymatgen|, an open source python library for material analysis\cite{ong2013python}, the package cannot handle any incommensurate structure nor commensurate structure of alloys. We used \verb|Pymatgen|'s code as the base and wrote our own code to parsed all structure in the data based. We also created visualization code to double check our code by comparing the parsed structures with both MAGNDATA online visualizer as well as the original publications of all materials.

\subsection{Computational Environment}
We implemented our models in Python 3.11.9 and trained them on our graphics processing unit cluster with CUDA version 11.4. To facilitate the model implementation and training, we used following important Python modules: \verb|Pymatgen| and \verb|ase| for handling material chemical structure \cite{larsen2017}, \verb|PyTorch| for managing the model training framework \cite{ansel2024}, \verb|PyTorch Geometric| for managing graph neural network models \cite{fey2025}, \verb|e3nn| for implementing our neural network models in a form that is equivariant for the Euclidean group ($E(3)$) \cite{geiger2022e3nn}, \verb|numpy| \cite{harris2020array}, \verb|scipy| \cite{virtanen2020} and \verb|pandas| \cite{mckinney2010} for data preparation and analysis, \verb|mendeleev| \cite{mentel2014} for easy access to important atomic features, \verb|matplotlib| \cite{hunter2007} and \verb|PyVista| \cite{sullivan2019pyvista} for visualization, and \verb|cmaes| for models' hyperparameter optimization \cite{nomura2026cmaes}.

\subsection{Propagation Vector Encoding}
The non-uniqueness of some predicting parameters can cause the prediction quality to degrade since the regressive nature of the neural network models would try to average all possible outputs. Hence, it is important to decide on the representations of these outputs to be either unique or invariant with all possible choices of the outputs.

From Eq.~(\ref{eq:modulation_site_cc}), the possible space for $\mathbf{k}$ is cut in half due to the deterministic connection between the Fourier components between the terms corresponding to propagation vectors and their negatives. On top of that, since the systems under consideration are all periodic and $\mathbf{k}$ are in the unit of reciprocal lattice vectors. The only viable $\mathbf{k}$-space that we want the model to learn is half of $[-0.5, 0.5)^3$ with, for any $\mathbf{k}\in [-0.5, 0.5)^3$, it is identified with $-\mathbf{k}$.

The choice for the half space is actually arbitrary, but we want the space to be some what symmetric to the permutation of the order of reciprocal lattice vector. Furthermore, for convenience, we also want to only deal with non negative value. Hence, we decided to map the $\mathbf{k}$'s domain from $[-0.5, 0.5)^3$ to $[0, 1)^3$ by modulo each component of $\mathbf{k}$ and only choose the points $\mathbf{k}$ with their component summation at most $3/2$. This can be done with out affecting the Fourier components due to the periodic nature of the systems. There are a few exception cases for this choice of half space. If the propagation vector has only 1 non-zero component, that component value must be at most $1/2$, this is because the negation of these vectors (after modulo back to $[0, 1)^3$) are also in the same half space.

\subsection{Fourier-Component Representation}
From Eq.(~\ref{eq:modulation_site}), we can see that the Fourier representation for the magnetic moment at each site is unique up to the choice of primitive atomic lattice origin. For example, if we decided to relabel the lattice points such that $\mathbf{n}\rightarrow\mathbf{n}+\mathbf{d}$ where $\mathbf{d}\in\mathbb{Z}^3$, to preserve the original structure, the magnetic moment under the new representation (decorated by prime) must obey
\begin{align*}
    \mathbf{m}'_{\alpha,\mathbf{n}+\mathbf{d}} &= \sum_\mathbf{k}\widetilde{\mathbf{m}}_{\alpha,\mathbf{k}}\exp(2\pi i\mathbf{k}\mathbf{n}^\top)\\
    \mathbf{m}'_{\alpha,\mathbf{n}} &= \sum_\mathbf{k}\widetilde{\mathbf{m}}_{\alpha,\mathbf{k}}\exp(-2\pi i\mathbf{k}\mathbf{d}^\top)\exp(2\pi i\mathbf{k}\mathbf{n}^\top).
\end{align*}
This implies that 
\begin{equation}\widetilde{\mathbf{m}}'_{\alpha,\mathbf{k}}=\widetilde{\mathbf{m}}_{\alpha,\mathbf{k}}\exp(-2\pi i\mathbf{k}\mathbf{d}^\top)
\end{equation}
which depending on the value of $\mathbf{k}$ the number of variations can be large for commensurate structures with small propagation vectors or even infinite for incommensurate structures. 

Unlike in the propagation vector case, there is no absolute origin for $\mathbf{n}$, we cannot directly use the trick of limiting the representation space especially for the incommensurate structure. Instead, we will just use the standard representation, and need to design the loss function for dealing with all varieties of unit cell selection which we will discussed in more detail in the loss function and training section.

\subsection{Equivariant Message Passing}
Given the limited size of the MAGNDATA dataset and the expected E(3)-equivariance of magnetic structures under Euclidean transformations of the input atomic structure, using an E(3)-equivariant graph neural network is highly advantageous. Such models naturally preserve rotational, translational, and reflection symmetries, eliminating the need for extensive data augmentation while enabling the network to learn physically consistent representations. To implement this, the model is built using the \verb|e3nn| framework, which provides operations that preserve equivariance throughout the message-passing process.

The core mathematical objects supported by \verb|e3nn| are irreducible representations of E(3), including scalars, vectors, pseudo vectors, and higher-order spherical harmonic components, along with their tensor products and decompositions. This allows conventional scalar message passing to be generalized to non-scalar features by replacing the learned edge kernel with a spherical harmonic expansion. The resulting equivariant message-passing update is given by
\begin{align*}
    h^{n+1}_j &= c_jf^n_j\otimes \sum_{\ell,m}R_\ell(\|\mathbf{r}_{j,i}\|_2)Y_{\ell,m}(\hat{\mathbf{r}}_{j,i}) \\
    f^{n+1}_i &= \sigma_\text{gate}\left(\sum_{j;(j,i)\in\mathcal{E}}h^{n+1}_j\right).
\end{align*}
Here, $f^n_j$ represent the feature of node $j$ at the layer $n$ of the model, $\otimes$ denotes the tensor product, $R_\ell$ is a learnable radial function depending only on the inter-node distance $\|\mathbf{r}_{j,i}\|_2$, $Y_{\ell,m}$ are spherical harmonic functions of the normalized relative direction $\hat{\mathbf{r}}_{j,i}$, and $\sigma_\text{gate}$ is an equivariant activation function provided by \verb|e3nn|. Since both $c_j$ and $R_\ell$ are scalar and therefore invariant under $E(3)$ while the spherical harmonic terms transform according to known equivariant rules, the entire message-passing operation remains equivariant as long as the node features are represented using compatible irreducible representations.

\subsection{Training Objectives}
In total, there are 5 targets our models need to predict: magnetic atom identification, number of non-zero propagation vectors, the value of those vectors in the unit of reciprocal lattice vectors, total magnetic moment amplitude of each atom, and the distribution as well as relative angles and phases of the magnetic moment Fourier components. To properly train the models for each target, the suitable loss functions are needed.

\subsubsection{Classification Targets}
Among the targets, three of them will be using classification style of prediction. these are the magnetic atom identification, number of non-zero propagation vectors, and the commensurate parts of the propagation vectors (one of the class is a dummy where any incommensurate parts and commensurate parts that are not one of the predefined classes will be put into). The loss function for classification tasks will be cross entropy loss with different number of classes depending on the prediction targets.

\subsubsection{Regression Targets}
There are two normal regression tasks in this works which are the prediction of incommensurate parts of the propagation vectors, and total magnetic moment amplitude of each atom. The loss function used for this prediction is mean square error loss.

\subsubsection{Fourier-Component Alignment Loss}
As mentioned in the result section, the global phase shift when the lattice got relabeled cause the non-uniqueness in representation for the magnetic moment Fourier components. This problem has different behavior depending on whether we are considering commensurate or incommensurate structure. There are finite number of variations that the phase shift can be in the commensurate structure while, for incommensurate, the choices are infinite. In fact, the variations for incommensurate structure is dense, i.e., one can find $\mathbf{d}$ that make the phase arbitrarily close to any phase.

Furthermore, we also assume that all magnetic structures considered in this work are of Heisenberg type, meaning that only the relative orientations between spins affect the magnetic structure. Under this assumption, all co-rotation of all magnetic moments in the structures are considered to be equivalent.

With these variation, we design the loss function for the distribution, relative angles and phases of the magnetic moment Fourier components such that, for any prediction and target pair, the model will perform alternate optimization between the global phase and global rotation, and use the lowest loss (mean square error) of the optimum phase and rotation as the prediction loss.

\subsection{Magnetic Structure Reconstruction from Model Outputs}
Our models, for each structure, predict 5 targets that needed to be post process to reconstruct the PMSR. Starting with propagation vectors prediction, the model will predict a fixed amount of vectors of which the components are, first, filled by looking up the classifier for commensurate components. If the component is classified into a dummy class, we will use the result from the regressor since it is either an incommensurate or small commensurate components. The final vectors will be filtered for only the first few vectors with the amount determined by the classifier for number of non-zero propagation vectors.

As for magnetic moment Fourier components, we first filter out the same amount of components as the propagation vectors (need to add one for the constant term). Next, we normalized the components with the prediction of total magnetic moment amplitude of each atom. Lastly, we masked out the non-magnetic atoms' to have zero magnetic moments with the magnetic atom identification classifier. 

After we got both components, if the task is to predict the magnetic structure, we can simply output the resulting PMSR. On the other hand, if the task is to compare the prediction and the known target, due to the phase and rotational freedom, an extra search for a proper global phase and rotation is required.

\section*{Acknowledgment}
AC acknowledges support from U.S. Department of Energy (DOE), Office of Science (SC), Basic Energy Sciences Award No. DE-SC0020148. RO thanks support from National Science Foundation (NSF) ITE-2345084. This research used resources of the National Energy Research Scientific Computing Center (NERSC), a Department of Energy User Facility using NERSC award DDR-ERCAP0037478 and BES-ERCAP0038195. AC, RO, ER, DP, YC, JJT, and ML thank the support from DOE
Genesis Mission Multimodal AI for 2D Quantum Magnets (MAIQMag) project. ML acknowledges the support from With support from the Future Energy Systems Center (FUEC) through the MIT Energy Initiative and the support from R. Wachnik.

\bibliographystyle{apsrev4-2}
\bibliography{references}

\end{document}